# Real and marginal isotope effects in cuprate superconductors

A. R. Bishop[1], A. Bussmann-Holder[2], O. V. Dolgov[2], A. Furrer[3], H. Kamimura[4], H. Keller[5], R. Khasanov[5], R. K. Kremer[2], D. Manske[6], K. A. Müller[5], A. Simon[2]

1) Los Alamos National Laboratory, Los Alamos NM87545, USA
2) Max-Planck-Institute for Solid State Research, Heisenbergstr. 1, D-70569 Stuttgart, Germany
3) Laboratory for Neutron Scattering, ETH Zürich & PSI Villigen, CH-5232 Villigen PSI, Switzerland
4) Department of Applied Physics, Faculty of Science, Tokyo University of Science, 1-3 Kagurazaka, Shinjuku-ku, Tokyo, Japan 162-8601
5) Physik Institut der Universität Zürich, Winterthurerstr. 190, CH-8057 Zürich, Switzerland
6) Institut für Theoretische Physik ETH Zürich, Hönggerberg, CH-8093 Zürich, Switzerland

We critically review recent and earlier results on isotope effects in cuprate superconductors and emphasize that the sample preparation and the isotope exchange and back exchange are crucial in understanding and interpreting the data. Only extremely careful preparation techniques yield reliable results and permit differentiation between real isotope effects and marginal ones. The former are substantial and highlight the lattice vibrational importance in cuprate superconductors.

The search for the microscopic origin of superconductivity in the copper oxide compounds has initiated new experimental techniques with ultra high resolution limits and refined crystal growth methods. From the very beginning the conventional phonon mediated BCS pairing mechanism was mostly ignored and more exotic pairing interactions discussed for these complex materials. A key experiment was the isotope effect on $T_c$ which is vanishing at optimum doping, but increases systematically with decreasing doping level to be maximum at the border to the antiferromagnetic state [1]. Most remarkably, this trend is universal to all copper oxide superconductors (see Figure 1).

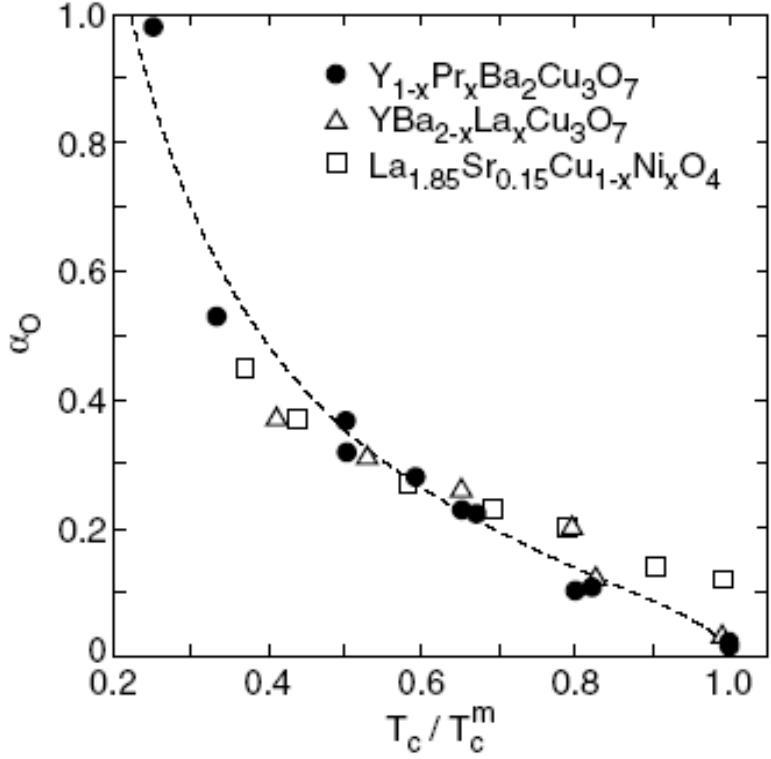

**Figure 1** Oxygen isotope effect exponent $\alpha_O$ versus $\overline{T}_c = T_c / T_c^m$ for various families of cuprate superconductors ($T_c^m$ corresponds to the maximum $T_c$ for a given family). Full circles: $Y_{1-x}Pr_xBa_2Cu_3O_{7-\delta}$ [1, 2, 3, 4] ; open triangles: $YBa_{2-x}La_xCu_3O_7$ [5] ; open squares : $La_{1.85}Sr_{0.15}Cu_{1-x}Ni_xO_4$ [6]. The dashed line corresponds to $\alpha_O = 0.25\sqrt{(1-\overline{T}_c)/\overline{T}_c}$ [7] (taken from [1]..

Soon after these reports on the doping dependent isotope effect, the question of which oxygen ions most contribute to it, was resolved experimentally by exchanging the isotope site selectively in $YBa_2Cu_3O_{7-\delta}$ [1, 2, 8, 9]. Here a two step isotope exchange process was used where simultaneously samples with complete isotope exchange were prepared together with those where either only the oxygen ions in planes, or in chains, e.g. apical, have been isotope substituted. The latter samples were prepared by using different annealing temperatures. In order to control the reliability of the site selective exchange, Raman measurements have been performed from which the respective lattice modes have been identified and the successful isotope exchange confirmed (Figure 2).

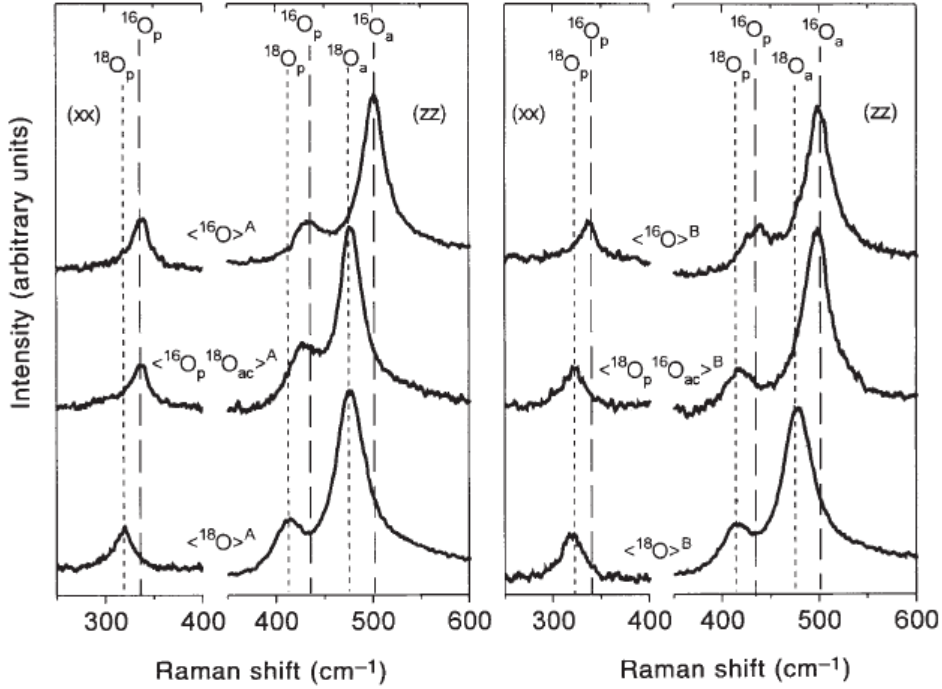

**Figure 2** Room temperature Raman spectra of the oxygen site-selectively substituted $YBa_2Cu_3O_{7-\delta}$ samples: triplet A (left), triplet B (right). For the (xx) polarization the line corresponds to the out-of-phase motion of the O(2,3) oxygen ions in the planes ($^{16}$O: 337 cm$^{-1}$, $^{18}$O: 320 cm$^{-1}$) and for the (zz) polarization the two lines correspond to the in-phase motion of the O(2,3) oxygen ions ($^{16}$O: 436 cm$^{-1}$, $^{18}$O: 415 cm$^{-1}$) and the stretching mode of the apical oxygen ion O(4) ($^{16}$O: 502 cm$^{-1}$, $^{18}$O: 478 cm$^{-1}$) [9].

Interestingly, the data revealed that the apical (chain) oxygen ions <u>do not</u> contribute to the isotope effect on $T_c$, but that it is carried only by the oxygen ions in the planes [2, 8, 9]. This observation excludes models where superconductivity is related exclusively to the apical oxygen ions. It is important to note, that the experiments were performed on the same batches of a sample and that a careful control of the data was performed by back exchanging the oxygen isotopes for the <u>same</u> samples in order to avoid different doping levels with different $T_c$'s introducing marginal effects.

Besides the doping-dependent and site-selective isotope effect on $T_c$ a striking isotope effect on the pseudogap formation temperature T* has been reported more recently by two different experimental techniques, EXAFS and inelastic neutron scattering, using the well defined cuprate superconductors $La_{1-x}Sr_xCuO_4$ and $HoBa_2Cu_4O_8$ [10, 11, 12]. Here the isotope effect is giant and, most interestingly, with sign reversed, i.e., the $^{18}$O samples show a drastically <u>enhanced</u> T* as compared to the $^{16}$O ones (Figure 3). In addition, a copper isotope effect also exists in the HoBa2Cu4O8 system for which the isotope coefficient is even larger than the one reported for oxygen [12]. Clearly, these results evidence that the lattice, i.e., the vibrational degrees of freedom, play a decisive role in the formation of the pseudogap phase and contributes substantially to the complexity of the phase diagram.

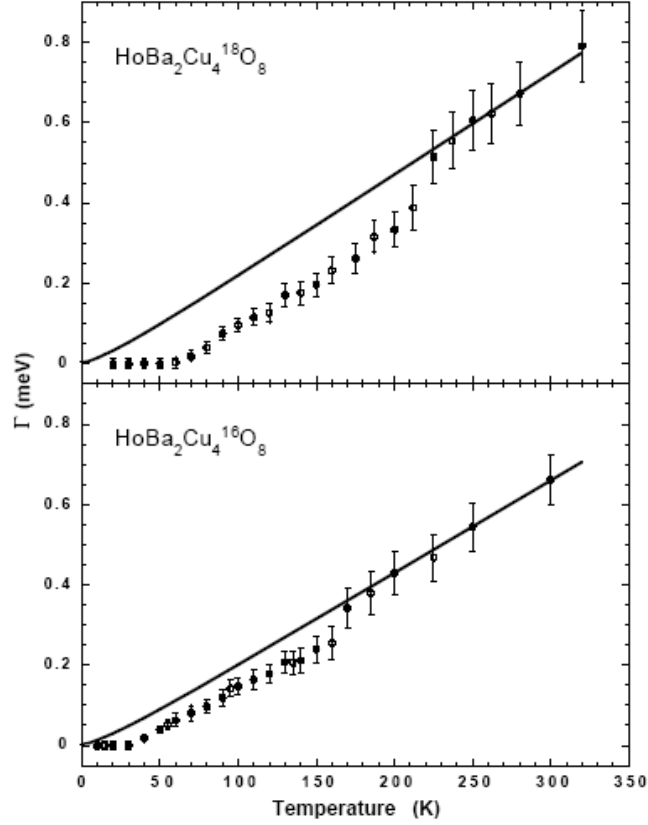

**Figure 3** Temperature dependence of the intrinsic line width $\Gamma$ (HWHM) corresponding to the $\Gamma_3 \rightarrow \Gamma_4$ ground state crystal field transition in $HoBa_2Cu_4{}^{16}O_8$ and $HoBa_2Cu_4{}^{18}O_8$ [11, 12]. The lines denote the line width in the normal state calculated from the Korringa law.

Another unusual isotope effect is that observed on the London penetration depth $\lambda_L$[13, 14, 15]. At optimum doping the isotope effect on $\lambda_L$ is substantial, although the one on $T_c$ is rather small (see Figure 4). With decreasing doping the isotope shift on $\lambda_L$ increases in a similar manner as the one on $T_c$ (see Figure 1). In addition, the isotope shifts of $T_c$ and $\lambda_L$ are correlated generically for various cuprates [1]. Note, that in a conventional BCS superconductor there exists no isotope effect on the penetration depth. Also in $MgB_2$ which is a two-gap superconductor mediated by electron-phonon interactions, no isotope effect on $\lambda_L$ exists [16]. Thus, these observations again emphasize the crucial role played by the lattice in the cuprates

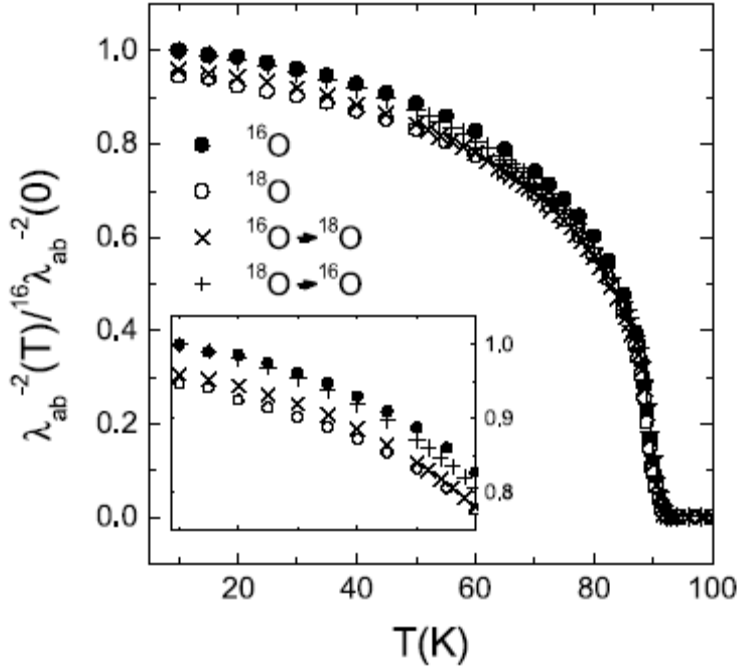

**Figure 4** Temperature dependence of the in plane penetration depth $\lambda_{ab}^{-2}$ normalized by $^{16}\lambda_{ab}^{-2}(0)$ for $^{16}$O- and $^{18}$O-substituted YBa$_2$Cu$_3$O$_{7-\delta}$ fine powder samples near optimal doping as obtained from low-field SQUID magnetization measurements. The inset shows the low temperature region between 10K and 60K. The reproducibility of the oxygen exchange procedure was checked by back exchange (crosses) [15].

Recently, an unexpected and not understood isotope effect was observed by ARPES techniques where a momentum and energy influence of the isotope substitution on the electronic binding energies has been detected [17]. These data have been taken on optimally doped Bi$_2$Sr$_2$CaCu$_2$O$_{8+\delta}$ samples. The data show reproducible and strong isotope effects which mainly appear in the broad high-energy humps. The strength of the effects depends strongly on the measuring momentum, and the magnitude of this effect closely correlates with the pair binding energy. Here also back exchange experiments have been performed in order to guarantee that the doping level is the same for both isotopes and that ambiguity in the interpretation of the data is removed.

Very recently STM data on Bi$_2$Sr$_2$CaCu$_2$O$_{8+\delta}$ surfaces have been reanalyzed [18]. From the indirectly obtained $d^2I/dV^2$ data a bosonic mode with energy of 52 meV was extracted and identified as a lattice mode since an oxygen isotope effect on it was observed. In addition, also an isotope effect on the superconducting energy gap was detected and correlated with the one on the bosonic mode. Bismuth oxocuprates are ill-defined even in the bulk due to incommensurate structural modulations which are not understood quantitatively. The samples of $Bi_2Sr_2CaCu_2O_{8+\delta}$ used for these STM experiments [18] are in this respect extremely poorly characterized since the $^{16}$O $T_c$ is 76K, whereas the $^{18}$O $T_c$ is 88K. However, in this regime of $T_c$ the isotope effect is nearly vanishing, and we conclude that these samples have a very different doping level, so that also the corresponding superconducting gaps are very different. In addition, it has been shown that most of the relevant lattice modes are doping dependent [18]. This in turn means that in two samples with different doping level a shift of a lattice mode by 6% is not due to the isotope exchange but due to the change in doping. A real isotope effect can only be inferred if the same sample is used in the experiment and the results are confirmed by back exchange.

To conclude, many isotope experiments have been performed on high $T_c$ copper oxide superconductors. Key to these experiments is that the isotope effect has always been measured on the same sample or the same batch of the sample in order to guarantee that the doping level is unchanged with the isotope exchange. Also, there have always been back exchange experiments to confirm the initial data. Since isotope effects result from electron-lattice interactions, their observation suggests that these are also important in understanding the physics of high $T_c$ compounds and highlight the vibronic character of the ground state. It is, however, impossible to conclude such effects if the doping levels of the two different isotope samples are different. Then, no conclusions can be drawn about the role played by the lattice since the doping level sensitively controls $T_c$, the gap value, the phonon energy, the pseudo gap etc. in very pronounced and well documented ways.

**Acknowledgement:** It is a pleasure to acknowledge many valuable discussions with M. Cardona and V. Z. Kresin.